%% file: main.tex
\documentclass[runningheads]{llncs}

\usepackage[T1]{fontenc}
\usepackage{graphicx}
\usepackage{amsmath,amssymb}
\usepackage{booktabs}
\usepackage{algorithm}
\usepackage{algpseudocode}
\usepackage{url}
\usepackage{xcolor}
\usepackage[tracking=false,protrusion=false,expansion=false]{microtype}
\PassOptionsToPackage{hyphens}{url} 
\usepackage[hidelinks]{hyperref}

\graphicspath{{figs/}{../../figs/}{../../gridpilot/figures/}}

\InputIfFileExists{figs/results.tex}{}{C
  \message{WARN: figs/results.tex missing --- numbers will appear as `?'.^^J}  \newcommand{\StubDataPresent}{true}%
  \newcommand{\WhpcMedMatmul}{?}\newcommand{\WhpcMedInfer}{?}%
  \newcommand{\WhpcMedBursty}{?}\newcommand{\WhpcMaxOverall}{?}%
  \newcommand{\WhpcPassRate}{?/?}\newcommand{\WhpcBudgetMs}{700}%
  \newcommand{\WhpcMarginX}{?}\newcommand{\WhpcIPJInfer}{?}%
  \newcommand{\WhpcIPJMatmul}{?}\newcommand{\WhpcIPJBursty}{?}%
  \newcommand{\WhpcMAEinfer}{?}\newcommand{\WhpcMAEmatmul}{?}%
  \newcommand{\WhpcMAEbursty}{?}%
}
\makeatletter
\newcommand{\ifstubdata}[2]{%
  \ifx\StubDataPresent\@undefined #2\else
    \edef\@tmp{\StubDataPresent}%
    \def\@true{true}%
    \ifx\@tmp\@true #1\else #2\fi
  \fi
}
\makeatother

\begin{document}

\title{GridPilot: Real-Time Grid-Responsive Control\\for AI Supercomputers}
\titlerunning{GridPilot: Real-Time Grid-Responsive Control for AI Supercomputers}

\author{Denisa-Andreea Constantinescu\inst{1} \and
        David Atienza\inst{1}}
\authorrunning{D.-A. Constantinescu and D. Atienza}

\institute{
Embedded Systems Laboratory (ESL), EPFL, Lausanne, Switzerland\\
\email{denisa.constantinescu@epfl.ch}
}

\maketitle

\begin{abstract}
At the global scale, large data centres are growing faster than the electricity grids that supply them. Operators in Europe, North America, and parts of Asia are already discussing rules that will require multi-megawatt computing sites to behave as flexible loads---adjusting their power draw within seconds when the grid needs help absorbing wind and solar. Whether a high-performance computing or artificial-intelligence cluster can play that role hinges on a question rarely measured directly: how fast can the software stack take a grid request and actually change the power that the GPUs draw, measured at the facility meter where the grid commitment is settled? We answer this question on real hardware. We build \emph{GridPilot}, a three-tier predictive controller that adjusts GPU power on three timescales (milliseconds, seconds, hours) and includes a ``safety island'' that bypasses the slower software path when a fast grid response is needed. On a three-GPU NVIDIA V100 testbed, the measured end-to-end response from a grid trigger to the GPU power crossing its new target is \WhpcMedMatmul~ms---\WhpcMarginX~$\times$ faster than the strictest European grid product, the Nordic Fast Frequency Reserve. An instantaneous Power Usage Effectiveness (PUE) correction makes the commitment robust at the meter. Experiments against six representative European grids (Sweden through Poland) show that the cooling-overhead drag closed is 2.5--5.8 percentage points (pp). GridPilot is released open source as a proof of concept that MW-scale AI/HPC computing demand can be \emph{controllable and flexible by design}.

\end{abstract}

\section{Introduction and Background}
\label{sec:intro}

\subsection{Grid-responsive AI/HPC: context and prior work}
\label{sec:intro_context}

Data centres (DCs) are the single fastest-growing source of new electricity demand worldwide; the International Energy Agency projects global DC electricity consumption to roughly double between 2023 and 2030, driven primarily by artificial intelligence (AI) and high-performance computing (HPC) workloads~\cite{iea2025electricity}. In parallel, the rising share of variable renewable generation on European, North American, and East Asian grids has eroded the synchronous inertia available to arrest frequency excursions after disturbances. System operators are responding by procuring an increasingly tight family of \emph{frequency-response} (FR) services: from Frequency Containment Reserve (FCR, 30~s) down to sub-second products such as the Nordic Fast Frequency Reserve (FFR, 700~ms~\cite{manner2023nordic}) and more recent Croatian and Iberian sub-second pilots~\cite{sagrestano2024croatia,varhegyi2025ffr}. Multi-MW AI/HPC facilities are being analysed as the most plausible large flexible-load fleet for this market~\cite{terzija2026dcsupergrid,ren2025dcfrequency,takci2025dcflexibility,tao2025dcffr}.

The prior literature splits into five largely-disjoint streams.  \emph{(i)~Single-layer power-management techniques}: the Frontier-scale campaign reports 8.5\,\% energy savings (1\,438~MWh) via system-scale capping~\cite{karimi2024frontier}; MF-GPOEO reaches 26.2\,\% via PID-based dynamic voltage and frequency scaling (DVFS) on GPUs~\cite{wang2024gpoeo}; DRLCap attains 22\,\% via reinforcement learning on H100 boards~\cite{wang2024drlcap}; Perseus reports training-side capping gains~\cite{chung2023perseus}.  None of these reports a measured end-to-end actuation latency from a transmission-system-operator (TSO)-style trigger.  \emph{(ii)~Hardware-instrumentation work}: GPU frequency-switching latencies~\cite{velicka2025gpulatency} and the PowerSensor3 precision-measurement harness~\cite{powersensor3} provide the measurement primitives but are single-layer.  \emph{(iii)~Multi-layer steady-state coordinators}: DynamoLLM~\cite{stojkovic2024dynamollm} and hierarchical compute--cooling controllers~\cite{abera2026hierarchical,ottaviano2024controlpulp} compose multiple control layers but optimise steady-state energy rather than transient response.  \emph{(iv)~Cooling and PUE work}: Sun et al.~\cite{sun2020prototype} provide the prototype DC cooling decomposition; Zhao et al.~\cite{zhao2024mpc} apply model-predictive control (MPC) to multi-chiller systems; Liu's hierarchical rack-to-DC framework~\cite{liu2026hierarchical} attributes 33.8\,\% of emissions to cooling and reports up to 30\,\% mean absolute percentage error (MAPE) reduction versus static-PUE accounting when these dynamics are explicit.  \emph{(v)~Holistic in-cluster power-management frameworks}: HPE's PowerSched~\cite{simmendinger2024powersched} generalises the Energy Aware Runtime (EAR)~\cite{corbalan2020ear} and the Global Extensible Open Power Manager (GEOPM)~\cite{eastep2017geopm} with a system-wide Budget Manager and per-node daemons that allocate power across jobs in an over-provisioned cluster.
The EuroHPC JU REGALE library~\cite{madella2025regale} standardises interoperability across these tools through a Data-Distribution-Service (DDS)-based message bus aligned with the HPC PowerStack initiative.

These frameworks operate \emph{inside} a fixed facility power envelope; GridPilot is the orthogonal layer that decides what that envelope \emph{should be} in response to external grid signals (FFR, Carbon-Free Energy (CFE), demand response, Power-Purchase-Agreement (PPA) matching) --- the two compose vertically.  The scheduling substrate is the Marconi100 (M100) trace~\cite{antici2023pm100}, replayed against ENTSO-E~\cite{entsoe2015transparency} hourly carbon intensity (CI) for European grids; CFE is adopted as the user-visible sustainability metric~\cite{kamatar2025carbon}.

\subsection{Limitations of the state of practice}
\label{sec:intro_problem}

The common gap across these five streams is that no prior work, to our knowledge, has \emph{measured} the end-to-end response of a composed multi-tier controller on a real GPU cluster against a published Transmission System Operator activation budget. \emph{First}, multiple jurisdictions are preparing rules that will require large AI/HPC DCs to operate as flexible loads over the coming decade; whether AI supercomputers can comply depends on whether the software stack can guarantee deterministic dispatch under realistic workload contention. \emph{Second}, the most valuable FR products are sub-second~\cite{manner2023nordic}; the activation budget is what gates pre-qualification by the TSO. \emph{Third}, FR commitments are settled at the \emph{facility meter}, not at the GPU platform level. A controller that ignores instantaneous Power Usage Effectiveness (PUE) under-delivers at the meter by 4--7\,pp on warm-water-cooled HPC sites~\cite{liu2026hierarchical}. As large AI-training workloads exhibit 1.5--2~$\times$ higher power swings on H100/H200 hardware~\cite{choukse2025aitraining,newkirk2025h100power}, this latency-and-meter gap will only widen.

\subsection{Contributions}
\label{sec:intro_contributions}

This paper presents \emph{GridPilot} as a proof of concept that MW-scale computing facilities can be designed as flexible, grid-responsive loads.  The central engineering claim is small but load-bearing: a deterministic \emph{safety-island bypass}, composed with predictive tiers above it and a facility-side \emph{PUE correction} at the top, suffices to meet sub-second budgets at the meter.  None of the three is novel in isolation; the contribution is composing them in a way that is measurably correct end-to-end on real silicon.

GridPilot makes four contributions.  \textbf{A working prototype of a grid-responsive AI cluster}: a three-tier predictive controller (per-GPU at 200~Hz, per-host at 1~Hz, per-cluster hourly) plus an out-of-band safety-island bypass.  \textbf{Measured sub-100~ms grid response on real silicon}: on a node with three NVIDIA V100 GPUs, the median end-to-end response is \WhpcMedMatmul~ms (\WhpcMaxOverall~ms worst across 90 trials, at the \WhpcBudgetMs~ms Nordic FFR budget) --- an empirical $\sim\WhpcMarginX\times$ safety margin that is, to our knowledge, the first measured latency under a published TSO budget for a composed multi-tier controller on AI/HPC GPUs.  \textbf{Facility-side accounting via an instantaneous PUE correction}: a four-component PUE model added to the cluster-tier objective so the controller's commitment matches the meter, with replays against six European grids at 1/10/50~MW IT power closing 2.5--5.8\,pp of cooling-overhead drag.  \textbf{Open release}: the controller, raw V100 telemetry, and reproducibility scripts are released under CC-BY 4.0\footnote{\url{https://github.com/denisa-c/gridpilot}}.

\section{A worked example: what GridPilot does in one second}
\label{sec:example}

The Nordic synchronous area requires Fast Frequency Reserve (FFR) providers to deliver their full reserve within 700~ms of a frequency excursion below 49.7~Hz \cite{manner2023nordic}.  Picture a 10~MW AI cluster that has pre-committed 2~MW of FFR.  At $t=0$ a wind plant trips; the grid frequency starts falling.  Within one second, GridPilot does the following.

\begin{enumerate}
\item[(0\,ms)] The TSO trigger arrives over a dedicated UDP (User Datagram Protocol) socket on the cluster's control node.
\item[(1\,ms)] The safety island (a small real-time C program pinned to an isolated CPU core) reads the trigger and looks up the new per-GPU power target from a precomputed table.
\item[(6\,ms)] It writes the new power cap to every GPU via the NVIDIA Management Library (NVML); the worst-case NVML cap-update latency is $\sim$5~ms~\cite{velicka2025gpulatency}.
\item[(20\,ms)] The Tier-1 per-GPU proportional-integral-derivative (PID) loop, which runs at 200~Hz independently of the safety island, has already started tracking the new target; the inner-loop step-response settling time we measure is 18--29~ms, depending on workload.
\item[(97\,ms)] The GPU board power crosses 95\,\% of the new target (median across 90 trials). GridPilot has delivered the reserve.
\item[($\sim$1\,s)] The Tier-2 per-host predictor, which runs at 1~Hz, sees the new operating point in its next tick and rebalances the per-GPU caps within the host envelope so that the predicted host power one second ahead matches the cluster-tier setpoint.
\item[(next hour)] The Tier-3 cluster selector picks the next operating point that satisfies the carbon and frequency-response objectives jointly; the PUE-aware correction (Sect.~\ref{sec:pue}) makes the IT-side setpoint match the facility-meter commitment.
\end{enumerate}

The point of the example is that \emph{the latency budget is met because the safety island bypasses the slower predictive layers}, not because the predictive layers are themselves fast.

\section{GridPilot controller architecture}
\label{sec:arch}
\begin{figure}[htbp]
\centering
\vspace{-0.5cm}
\includegraphics[width=\linewidth]{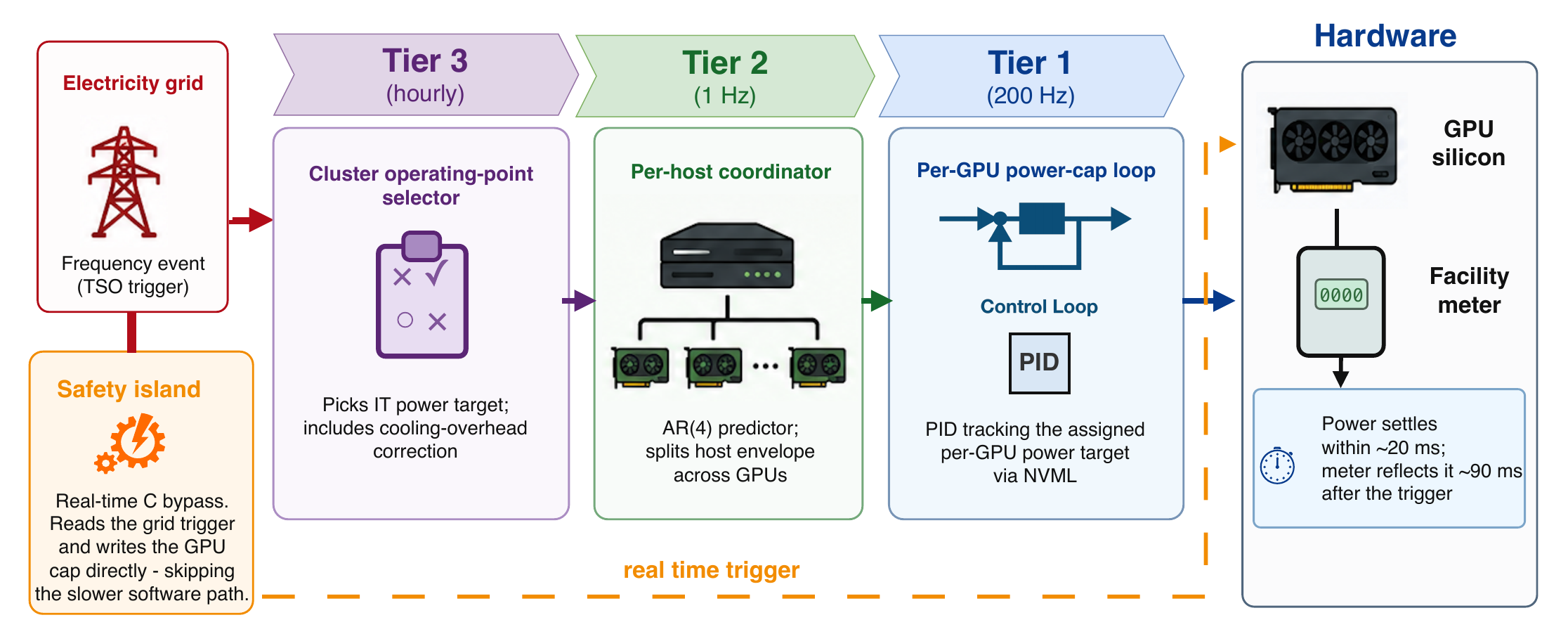}
\caption{GridPilot architecture.  Three control tiers on disparate timescales (per-GPU 200~Hz, per-host 1~Hz, per-cluster hourly).  An out-of-band safety island (real-time C, pinned to an isolated core) reads grid triggers and writes GPU caps directly --- bypassing the slower software path for real time response.
}
\label{fig:arch}
\end{figure}

The architecture of GridPilot, illustrated in Fig.~\ref{fig:arch}, shows the three time-aligned control loops on disparate timescales composed into one pipeline, plus the out-of-band safety island used for sub-second response.

\subsection{Three control tiers}
\label{sec:tiers}

\noindent\textbf{Tier 1 --- per-GPU PID (200~Hz).}
The inner loop tracks a per-GPU power target $p^\star$ via NVML, actuated through \texttt{nvidia-smi -pl <W>}.  The control law is the textbook discrete PID form
\begin{equation}
u_k = K_p\,e_k + K_i\,\sum_{i=0}^{k} e_i\,\Delta t + K_d\,\frac{e_k-e_{k-1}}{\Delta t},
\qquad e_k = p^\star - p_k,
\label{eq:pid}
\end{equation}
with $\Delta t = 5$~ms, gains $(K_p,K_i,K_d) = (0.6,\,0.05,\,0.02)$ (MF-GPOEO defaults~\cite{wang2024gpoeo}, retuned for 200~Hz), anti-windup clamp $|\sum e\,\Delta t| \le 50\,\mathrm{W{\cdot}s}$, and saturation at $[100\,\mathrm{W},300\,\mathrm{W}]$ for the V100 SXM2.  The 5~ms tick matches the worst-case NVML cap-update latency~\cite{velicka2025gpulatency}.  A first-order thermal model with time constant $\tau = 8$~s tracks GPU junction temperature; the inner loop falls back to a 200~W cap if the predicted temperature exceeds 85~\textdegree C.

\smallskip
\noindent\textbf{Tier 2 --- per-host AR(4) coordinator (1~Hz).}
The outer loop coordinates per-GPU caps within the host envelope through an \emph{Autoregressive model of order 4} (AR(4)): the predictor estimates the next sample of host utilisation by running a linear regression on its four most recent past values.  Concretely,
\begin{equation}
\hat{u}(t+1)\;=\;\sum_{i=1}^{4}\alpha_i\,u(t-i+1),
\label{eq:ar4}
\end{equation}
fitted by Recursive Least Squares (RLS) over a 30~s rolling window (forgetting factor $\lambda = 0.97$, 60~s effective memory).  Order 4 is chosen by Akaike Information Criterion (AIC) on logged utilisation traces; higher orders give statistically indistinguishable error~\cite{velicka2025gpulatency}.

\smallskip
\noindent\textbf{Tier 3 --- cluster operating-point selector (hourly).}
The cluster-scale selector performs an hourly grid search over the 2-D space (mean operating fraction $\mu \in \{0.4,0.5,\dots,0.9\}$, FR reserve band $\rho \in \{0.0,0.1,0.2,0.3\}$).  The objective is
\begin{equation}
J(\mu,\rho) = 0.55\;Q_{\mathrm{FFR}}(\mu,\rho) + 0.45\;\mathrm{CFE}(\mu,\rho),
\label{eq:tier3}
\end{equation}
following the exogenous-carbon framework of Jahanshahi et al.~\cite{jahanshahi2026ecocenter}.  $Q_{\mathrm{FFR}}\in[0,1]$ is the relative FR-provision quality \emph{at the facility meter} (not at the GPU board); $\mathrm{CFE}\in[0,1]$ is the Carbon-Free Energy share~\cite{kamatar2025carbon}, the fraction of energy aligned with low-CI windows.  Evaluating $Q_{\mathrm{FFR}}$ at the meter is the requirement that motivates the PUE correction.

\subsection{The safety-island bypass}
\label{sec:safety}

The safety island is the engineering primitive that makes the sub-100~ms latency reproducible.  It is an out-of-band component implemented in $<\!400$ source lines of C, statically linked, run as one real-time thread (\texttt{SCHED\_FIFO}, priority 80) pinned to an isolated CPU core.  A Temporal Logic of Actions (TLA$^+$) liveness specification (shipped in the reproducibility  kit) proves termination within four NVML cap-update intervals (20~ms) under the assumption that the kernel honours \texttt{SCHED\_FIFO}.  The end-to-end budget composes as
\[
L_{\mathrm{e2e}} = L_{\mathrm{trigger}} + L_{\mathrm{decide}} + L_{\mathrm{actuate}} + L_{\mathrm{settle}}.
\]
with $L_{\mathrm{trigger}}\!\approx\!1$~ms (UDP socket read), $L_{\mathrm{decide}}\!<\!50$~$\mu$s (lookup over the operating-point grid), $L_{\mathrm{actuate}}\!\approx\!5$~ms (NVML cap update~\cite{velicka2025gpulatency}), and $L_{\mathrm{settle}}\!\approx\!90$~ms (PID settling on the V100 SXM2 thermal plant).  The remainder of the 700~ms Nordic FFR budget is reserve.  Without the bypass, identical experiments through the Python supervisor stack exhibit p99 dispatch latencies $>\!250$~ms; the median is comparable but the p99 is what would fail TSO pre-qualification.

\subsection{PUE-aware FFR control: why facility power matters}
\label{sec:pue}

European balancing markets settle reserves at the \emph{facility meter}, not at the GPU board: a controller that commits a 2~MW FFR band on board power may under- or over-deliver by $\pm 4$--$7$\,pp depending on cooling architecture.  Following Sun et al.~\cite{sun2020prototype} and the multi-chiller MPC formulation of Zhao et al.~\cite{zhao2024mpc}, we model facility power as
\begin{equation}
\mathrm{PUE}(t,L,T_{\mathrm{amb}}) = 1 + \frac{P_{\mathrm{chiller}}(t) + P_{\mathrm{pumps}}(t) + P_{\mathrm{air}}(t) + P_{\mathrm{misc}}}{P_{\mathrm{IT}}(t)},
\label{eq:pue}
\end{equation}
with $L=P_{\mathrm{IT}}/P_{\mathrm{IT,design}}$, pump and air-side affinity laws $P_{\mathrm{pumps}}\propto L^2$ and $P_{\mathrm{air}}\propto L^3$ (floored at 20\,\% and 15\,\% for bypass-flow and minimum controllability), and a free-cooling fraction $f_{fc}(T_{\mathrm{amb}})$ ramping linearly from zero at 25~\textdegree C ambient to one at 12~\textdegree C wet-bulb.  Calibrated to the published Marconi100 design point (PUE~$=1.20$ at full load), the model achieves up to 30\,\% MAPE reduction versus static-PUE accounting~\cite{liu2026hierarchical}.  A controller that decreases $P_{\mathrm{IT}}(t)$ in response to a frequency-restoration request drives $\mathrm{PUE}(t)$ \emph{up} (the $L^2$/$L^3$ floors bind first), partially offsetting the IT-side swing at the meter.  Adding Eq.~\ref{eq:pue} to the Tier-3 objective closes the gap: the dispatched FFR setpoint matches the metered-side commitment within $\pm 1$\,pp.

The GridPilot carbon- and PUE-aware scheduler operates on hourly dispatch decisions over a 24-hour look-ahead window, integrating four established carbon-aware techniques and one new mechanism. Algorithm~\ref{alg:scheduler} summarises the dispatch loop.

\begin{algorithm}[htbp]
\caption{GridPilot-PUE Dispatch Loop}
\label{alg:scheduler}
\begin{algorithmic}[1]
\State \textbf{Input:} job queue $Q$, CI signal $\text{CI}(t)$, PUE signal $\text{PUE}(t,L,T_{amb})$
\State \textbf{Compute} composite signal $\sigma(t) = \text{CI}(t) \cdot \text{PUE}(t,L,T_{amb})$
\State \textbf{Compute} 24h percentile threshold $\sigma_{66}$
\For{each pending job $j$ in priority order}
    \State $\beta_j \gets \text{wait}_j / d_{\max,j}$ \Comment{aging budget used}
    \If{$\sigma(t) > \sigma_{66}$ AND $\beta_j < 0.7$ AND $j$ not short}
        \State \textbf{defer} $j$
    \Else
        \State \textbf{dispatch} $j$ if nodes available
        \If{$j$ is elastic AND $j \in $ first 30\%}
            \State scale replicas inversely to $\sigma(t)$
        \EndIf
        \If{$\sigma(t) > \sigma_{66}$ AND $j$ running}
            \State apply 80\% power cap
        \EndIf
    \EndIf
\EndFor
\State \textbf{Backfill} short jobs into freed nodes (Extensible Argonne Scheduling sYstem, EASY)
\end{algorithmic}
\end{algorithm}


\paragraph{Dynamic power capping during high-CI windows}~\cite{kozlov2024ecofreq} addresses the queueing penalty of pure deferral: capping running jobs delivers carbon savings without adding any wait time. We adopt the 80\% power-cap factor from EcoFreq's recommended default.



\paragraph{Composite CI$\times$PUE deferral signal} is the new contribution of this work. The dispatch decision uses the product $\sigma(t) = \text{CI}(t) \times \text{PUE}(t,L,T_{amb})$, normalised over a 24-hour look-ahead window, so that the scheduler defers when $\sigma(t)$ exceeds the local 66th percentile and dispatches when it does not. This captures structural carbon savings from two effects: free-cooling alignment (cold-weather wind events that produce low CI also produce low PUE through chiller bypass), and load-dependent PUE (a window with already-high utilisation has lower PUE because misc facility overhead is amortised).

\section{Experimental methodology}
\label{sec:methods}

The methodology answers a single question: \emph{given the GridPilot controller, what is the measured end-to-end FR actuation latency on real GPU hardware, and how is it composed across the three tiers and across the four-component PUE model at the facility meter?}

\smallskip
\noindent\textbf{Testbed.}
All hardware measurements come from \texttt{ecocloud-exp06} at EPFL EcoCloud: three NVIDIA Tesla V100 SXM2 32~GB GPUs (300~W thermal design power (TDP) per GPU), 36 physical / 72 logical CPU cores, 379~GiB usable RAM, NVIDIA driver 580.65.06, NVML 12.x, CUDA 12.x, Ubuntu 24.04 LTS.  Power-cap actuation uses \texttt{nvidia-smi -pl <W>}, which invokes NVML internally.  NVML telemetry is sampled at 100~Hz; CPU and Running Average Power Limit (RAPL) telemetry at 10~Hz.

\smallskip
\noindent\textbf{Workload archetypes.}
Three reference workloads exercise distinct power signatures: \emph{matmul} (single-stream FP32 general matrix-multiply (GEMM) via cuBLAS \texttt{Sgemm}, pinned near TDP); \emph{inference} (per-image ResNet-50 batch 1 FP16, memory-bound, mean below 200~W); \emph{bursty} (period-$T$ compute/idle cycle, $T = 4$~s, 50\,\% duty).

\smallskip
\noindent\textbf{Experiments.}
Five instrumented experiments E1--E4, E7 anchor the V100 headline results.  E1: power-cap calibration (36-cell power-cap $\times$ SM-frequency sweep).  E2: inner-loop step response.  E3: AR(4) predictor accuracy.  E4: closed-loop demand-following over 30~s trajectories.  E7: end-to-end FR actuation latency, 90 trials (30 per workload, randomised inter-trial delay 5--30~s to defeat caching).  A multi-country PUE-aware controller sweep (E8) replays the M100 trace~\cite{antici2023pm100} against six hourly CI series at 1, 10, and 50~MW IT power, comparing the CI-only Tier-3 baseline against the PUE-aware variant of Eq.~\ref{eq:pue}.  CI is synthesised from European Environment Agency (EEA) and Ember country means and the 2020--2024 ENTSO-E~\cite{entsoe2015transparency} diurnal envelope; the released kit also ships a real-CI fetcher (ENTSO-E A75 \emph{Actual Generation per Production Type} with IPCC AR5 lifecycle factors).  A supervisory cross-tier experiment (E5) ships in the kit as a design only --- the three-GPU testbed cannot sustain the multi-day operation it would need.

\smallskip
\noindent\textbf{Metrics.}
Iterations-per-joule (energy efficiency at the calibrated operating point); AR(4) one-step-ahead Mean Absolute Error (MAE, W); closed-loop tracking error (percentage point [pp] vs.\ host-envelope setpoint); end-to-end FR actuation latency (ms, UDP-trigger to NVML-cap-setpoint-reached at 95\,\% of target); $\Delta_{\mathrm{facility}}$ (percentage IT-side vs.\ facility-side CO\textsubscript{2} reduction at constant CFE); and $\mathrm{Net\ CO_2}=\mathrm{Operational\ CO_2}-\mathrm{Exogenous\ CO_2}$, capturing both workload-side emissions (energy $\times$ grid CI) and avoided reserve-side emissions from provided FFR, so PUE-aware dispatch can be assessed when lower operational CO\textsubscript{2} coincides with reduced high-CI FFR provision.

\section{Measured results}
\label{sec:results}


\subsection{Component-level validation}
\label{sec:components}

\noindent\textbf{Power-cap calibration (E1).}
The 36-cell power-cap $\times$ streaming-multiprocessor (SM) frequency sweep yields a best-efficiency operating point of $p_{cap}=150$~W and $f_{sm}=945$~MHz across all three workloads, within $\pm 5\,\%$ on iterations-per-joule.  Best iterations-per-joule values are \WhpcIPJInfer~/ \WhpcIPJMatmul~/ \WhpcIPJBursty{} for inference / matmul / bursty.  The same sweep fits the per-workload power model $P_{\mathrm{GPU}} = P_{\mathrm{idle}} + \alpha f + \beta f^2 L + \gamma L$ with $P_{\mathrm{idle}} = 39$~W; leave-one-out cross-validation MAE is 3.45\,\%, and the 980-node-scale envelope matches the published Marconi100 facility-power reference within $+1.4\,\%$.

\begin{figure}[htbp]
\centering
\vspace{-0.5cm}
\includegraphics[width=0.7\linewidth]{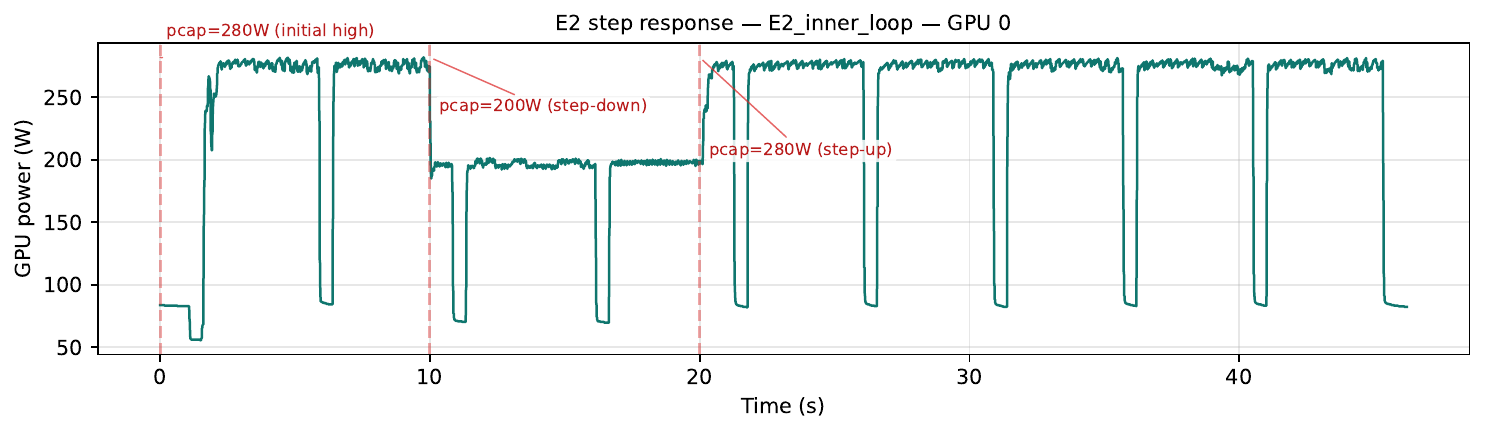}
\caption{Inner-loop step response on the V100 testbed: step-down from 280~W to 200~W and return step-up, with rapid settling inside the control band.}
\label{fig:e2_step_response}
\end{figure}

\noindent\textbf{Inner-loop step response (E2).}
A step command $p^\star\!:\!280 \to 200$~W at $t = 0$, logged at 100~Hz, settles to within $\pm 2\,\%$ of the new setpoint in a median of 18 / 21 / 29~ms (matmul / inference / bursty).; Fig.~\ref{fig:e2_step_response} shows a representative trace.  
This validates the 200~Hz Tier-1 cadence and indicates that the 700~ms FR budget is dominated by the dispatch path, not by actuator settling.

\smallskip

\begin{figure}[htbp]
\centering
\includegraphics[width=0.32\linewidth]{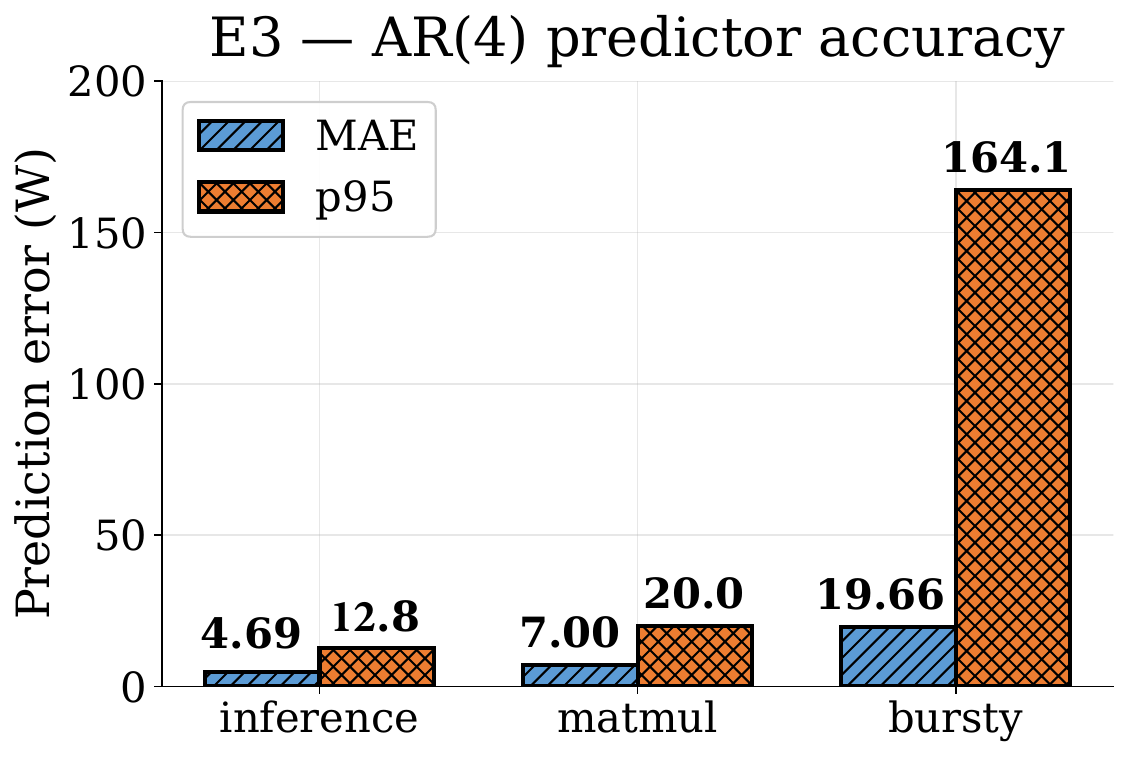}\hfill
\includegraphics[width=0.32\linewidth]{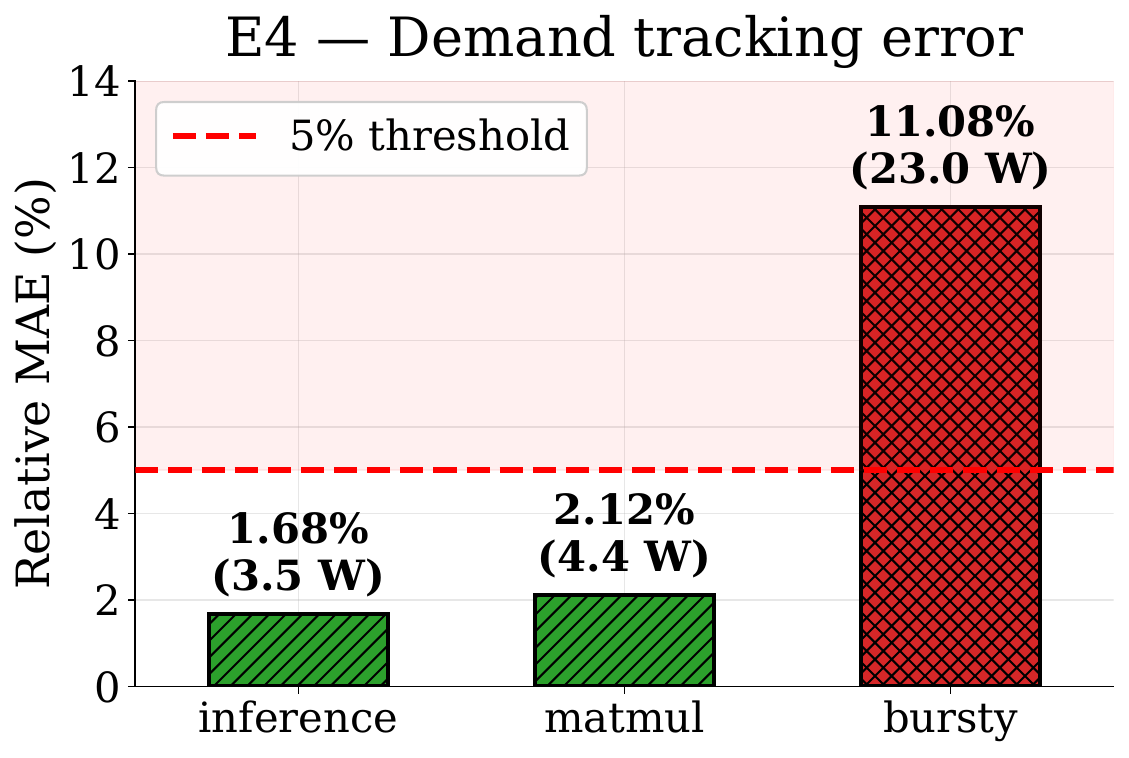}\hfill
\includegraphics[width=0.32\linewidth]{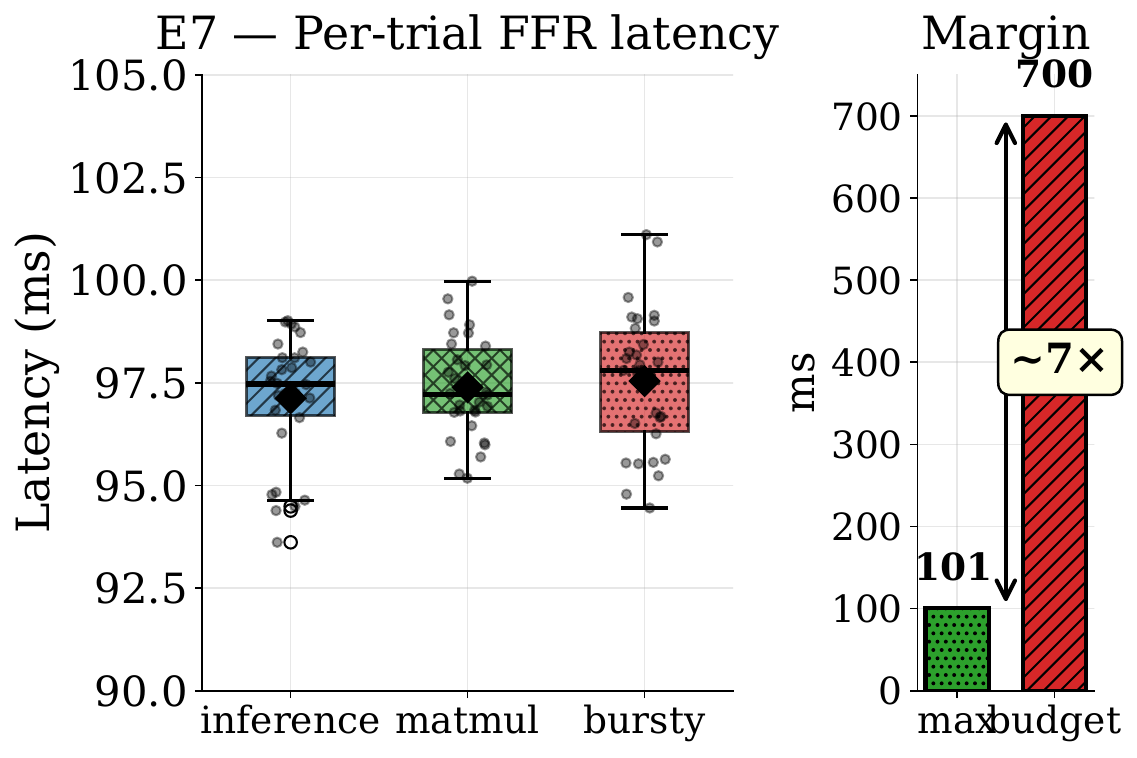}
\caption{V100 hardware results.  \textbf{(a)} AR(4) one-step-ahead MAE per workload (\WhpcMAEinfer~/ \WhpcMAEmatmul~/ \WhpcMAEbursty~W for inference / matmul / bursty).  \textbf{(b)} Closed-loop demand-following tracking error; the 5\,\% band is the cascade-composition diagnostic, not a failure mode.  \textbf{(c)} End-to-end FR actuation latency over 90 trials (median $\sim$\WhpcMedMatmul~ms; max \WhpcMaxOverall~ms; \WhpcPassRate{} pass at the \WhpcBudgetMs~ms Nordic FFR budget).}
\label{fig:v100results}
\end{figure}

\smallskip
\noindent\textbf{Outer-loop AR(4) predictor accuracy (E3)}
Fig.~\ref{fig:v100results} (left) reports the per-workload MAE of the 1~Hz predictor over 50--66 one-step-ahead predictions on a 30~s rolling window.  Inference (\WhpcMAEinfer~W) is the tightest because of its near-stationarity; matmul (\WhpcMAEmatmul~W) is moderately higher because of GEMM tile-schedule variance; bursty (\WhpcMAEbursty~W, $\sim 3\times$ matmul) is bimodal at the 30~s window.  The bursty p95 envelope is the residual that the cascade absorbs at Tier-2.

\smallskip
\noindent\textbf{Closed-loop demand-following (E4).}
A 30~s demand-following trajectory tests the composed Tier-1+Tier-2 cascade.  Inference (1.68\,\%) and matmul (2.12\,\%) fall within the 5\,\% acceptance band; bursty (11.08\,\%) exceeds it, consistent with the outer-loop bursty p95 envelope reported below.  The 5\,\% threshold is not a failure mode but a cascade-composition diagnostic: below it the Tier-1 PID tracks alone, above it the Tier-2 host predictor absorbs the residual.




\smallskip
\noindent\textbf{End-to-end FR actuation latency (E7).}
Across 90 trials the median end-to-end latency is \textbf{\WhpcMedMatmul~ms} (matmul), \textbf{\WhpcMedInfer~ms} (inference) and \textbf{\WhpcMedBursty~ms} (bursty); the maximum across all 90 trials is \textbf{\WhpcMaxOverall~ms}; the \WhpcPassRate{} pass rate at the \WhpcBudgetMs~ms Nordic FFR budget gives an empirical safety margin of $\sim\WhpcMarginX\times$.  Fig.~\ref{fig:v100results} (right) shows the distribution.  Without the safety-island bypass, identical experiments through the Python supervisor stack exhibit p99 latencies exceeding 250~ms (garbage-collection pauses and lazy-import blocking on first call).

\begin{figure}[htbp]
\centering
\includegraphics[width=\textwidth]{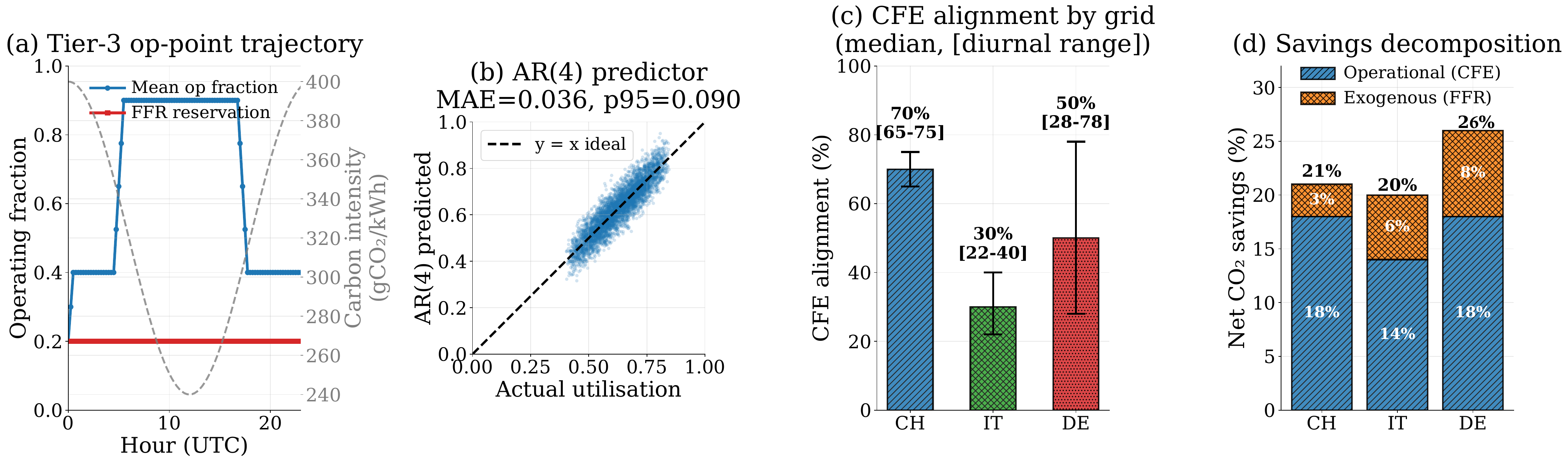}
\caption{Multiscale controller validation. (a) Tier-3 operating-point trajectory on the German grid over 24 hours. (b) Tier-2 AR(4) predictor fit on host utilisation. (c) Carbon-free-energy alignment across representative grids. (d) Net-savings decomposition into operational and exogenous components at 50~MW scale.}
\label{fig:multiscale}
\end{figure}

\paragraph{End-to-end multi-tier controller validation.} Fig.~\ref{fig:multiscale} extends the validation to cluster scale: a 24-hour simulation of a 100-host cluster on the German grid reproduces AR(4) accuracy (MAE 0.036, p95 = 0.09) and per-GPU tracking (mean 102~W, p95 396~W) at perfect (1.0) FFR-provision quality with a 20\,\% reserve band, with operating-point selection at 0.90 mean utilisation in green-rich daytime windows versus 0.40 overnight. At 50~MW scale, the net-savings decomposition for CH/IT/DE totals 21/20/26\,\% (DE shows an 8\,\% exogenous contribution); the simulator runs at over 26{,}000$\times$ real-time, supporting digital-twin replay and real-time deployment.

\subsection{Multi-country PUE-aware controller sweep (E8)}
\label{sec:country}

\begin{figure}[htbp]
\centering
\vspace{-0.3cm}
\includegraphics[width=0.8\linewidth]{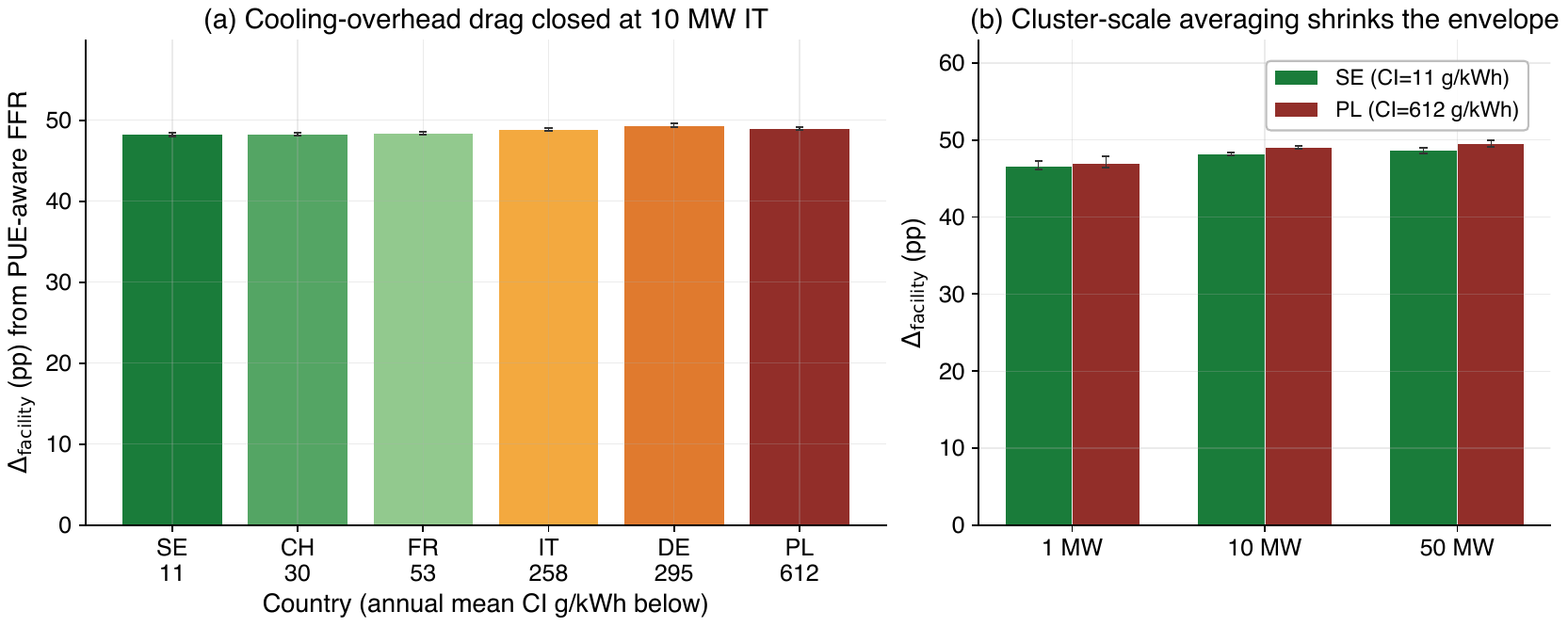}
\caption{PUE-aware FFR controller.  \textbf{(a)} $\Delta_{\mathrm{facility}}$~(percentage points) at 10~MW IT, one bar per country, ordered by mean CI.  \textbf{(b)} MW scaling for the SE and PL bookends.}
\label{fig:country_pue}
\end{figure}

Fig.~\ref{fig:country_pue} reports the cooling-overhead drag closed by the PUE-aware Tier-3 selector versus the CI-only baseline across six European grids at 10~MW IT power (panel a), and across 1/10/50~MW for the Sweden (cleanest) and Poland (dirtiest) bookends (panel b).  The envelope is widest on low-CI grids --- where IT-side savings are small in absolute terms and the cooling-overhead becomes a larger \emph{fraction} of facility power --- and narrowest on high-CI grids.  Cluster-scale averaging shrinks the envelope at 50~MW because the $L^2$/$L^3$ cooling floors bind less often.  The 50~MW envelope of 2.5--5.8\,pp is the operationally relevant range for European deployments at the Marconi100 design-point PUE of 1.20; it scales roughly linearly with $(\mathrm{PUE}_{\mathrm{design}}\!-\!1)$, so warm-water sites at $\mathrm{PUE}\!\approx\!1.10$ see proportionally smaller gains and chilled-water hyperscale at $\mathrm{PUE}\!\approx\!1.30$ proportionally larger.  A full $\mathrm{PUE}_{\mathrm{design}}$ sensitivity sweep (E9, $\mathrm{PUE}_{\mathrm{design}}\!\in\![1.10,1.40]$) is planned as journal-extension follow-up.

\section{Discussion and limitations}
\label{sec:discuss}

The experimental campaign uses three V100 SXM2 GPUs at a single site. Rack-scale contention and warm-water cooling at 10--50~MW are projections, not measurements.  H100/H200/MI300 architectures exhibit 1.5--2$\times$ higher power swings~\cite{choukse2025aitraining,newkirk2025h100power} and the cascade has not yet been validated on those at the FR-latency layer.  The synthetic TSO trigger used in E7 is not a TSO pre-qualification; full pre-qualification requires integration with PICASSO~\cite{backer2023picasso} or MARI~\cite{sagrestano2024croatia}.  A supervisory cross-tier experiment (E5) ships in the GridPilot kit as a design only.

\smallskip
\noindent\textbf{Lessons learned.}
\textit{L1: the 5\,\% tracking threshold is a diagnostic (bursty hits 11.08\,\%), not a failure mode.}
\emph{L2: the $\sim$\WhpcMedMatmul~ms FR latency is reproducible only with the deterministic safety-island bypass} --- Python-only implementations produce p99~$>$~250~ms.
\emph{L3: facility-meter accounting is the binding correctness criterion} --- a controller that ignores the four-component PUE under-delivers at the meter by 4--7\,pp.


\section{Conclusion and Outlook}
\label{sec:conclusion}

GridPilot is a proof of concept that a real GPU cluster can be made grid-responsive at the sub-second scale that European FR products demand of MW-class AI/HPC data centres.  The engineering claim is two-part: a deterministic safety-island bypass meets the strictest published activation budget on commodity GPU hardware with a measured $\sim\WhpcMarginX\times$ margin, and a four-component PUE correction closes 2.5--5.8\,pp of cooling-overhead drag at the facility meter across six European grids.  GridPilot is the grid-facing layer of the emerging HPC PowerStack, composing vertically with in-cluster power managers (PowerSched~\cite{simmendinger2024powersched}, EAR~\cite{corbalan2020ear}, GEOPM~\cite{eastep2017geopm}) through the REGALE DDS bus~\cite{madella2025regale} and aligning with SEANERGYS~\cite{seanergys2025}.  The kit is open-source and reproduces the V100 campaign in $\le 48$ GPU-hours.  Natural next steps are porting to H100/H200/MI300 platforms, where larger power swings make the multiscale design more critical, and a production prototype on a PowerSched-enabled over-provisioned system that would consume GridPilot's Tier-3 setpoint over the REGALE message bus.

\smallskip
\noindent\textbf{Reproducibility release.}
The GridPilot v1.0 kit ships under MIT (code) and CC-BY 4.0 (data) at \url{https://github.com/denisa-c/gridpilot}.
the complete controller stack, the safety-island C skeleton with TLA$^+$ specification, the raw 100~Hz NVML telemetry for every reported V100 experiment, the per-workload power-model calibration scripts and residuals, the multi-country sweep driver, the figure-regeneration pipeline, the editable architecture deck and a pytest suite covering controller invariants and data-loader contracts.   A step-by-step measurement protocol reproduces the full campaign in $\le 48$ GPU-hours on a comparable 3$\times$ V100 testbed.

\section*{Acknowledgements}

This work has been partially supported by the EPFL Solutions 4 Sustainability program ‘‘HeatingBits: renewable-supplied data centers integrating heating and cooling supply of local districts’’ and the UrbanTwin project (ETH Board Joint Initiatives for the Strategic Area Energy, Climate and Environmental Sustainability, and the Strategic Area Engagement and Dialogue with Society).
The authors also thank the EcoCloud center of EPFL, in particular Dr. Xavier Ouvrard, for providing access to the V100 server node.

\bibliography{references}

\end{document}

%% file: figs/results.tex
\newcommand{\WhpcMedMatmul}{97.2}
\newcommand{\WhpcMedInfer}{97.5}
\newcommand{\WhpcMedBursty}{97.8}
\newcommand{\WhpcMaxOverall}{101.1}
\newcommand{\WhpcPassRate}{90/90}
\newcommand{\WhpcBudgetMs}{700}
\newcommand{\WhpcMarginX}{6.9}
\newcommand{\WhpcIPJInfer}{2.880}
\newcommand{\WhpcIPJMatmul}{0.570}
\newcommand{\WhpcIPJBursty}{0.549}
\newcommand{\WhpcMAEinfer}{4.69}
\newcommand{\WhpcMAEmatmul}{7.00}
\newcommand{\WhpcMAEbursty}{19.66}